\documentclass[
 reprint,
superscriptaddress,
nofootinbib,
bibnotes,
 amsmath,amssymb,
 aps,
 prl,
floatfix,
]{revtex4-2}

\usepackage{lipsum} 
\usepackage{graphicx}% Include figure files
\usepackage{dcolumn}% Align table columns on decimal point
\usepackage{bm}% bold math
\usepackage{hyperref}% add hypertext capabilities
\begin{document}

%\preprint{APS/123-QED}

\title{Photoinduced dynamics of flat bands in the kagome metal CoSn}% Force line breaks with \\

\author{D.~Puntel}
%\email{denny.puntel@elettra.eu}
 \affiliation{Dipartimento di Fisica, Università degli Studi di Trieste, 34127 Trieste, Italy}%Lines break automatically or can be forced with \\

 \author{W.~Bronsch}
 \affiliation{Elettra - Sincrotrone Trieste S.C.p.A., Strada Statale 14, km 163.5, Trieste, Italy}

 \author{M.~Tuniz}
 \affiliation{Dipartimento di Fisica, Università degli Studi di Trieste, 34127 Trieste, Italy}
 
 \author{M.~Kang}
 \affiliation{Department of Physics, Massachusetts Institute of Technology, Cambridge, MA, USA}

 \author{P.~Neves}
 \affiliation{Department of Physics, Massachusetts Institute of Technology, Cambridge, MA, USA}
 
 \author{S.~Fang}
 \affiliation{Department of Physics, Massachusetts Institute of Technology, Cambridge, MA, USA}
 
 \author{E.~Kaxiras}
 \affiliation{Department of Physics, Harvard University, Cambridge, MA, USA}
 \affiliation{John A. Paulson School of Engineering and Applied Sciences, Harvard University, Cambridge, MA, USA}
 
 \author{J.~G.~Checkelsky}
 \affiliation{Department of Physics, Massachusetts Institute of Technology, Cambridge, MA, USA}

 \author{R.~Comin}
 \affiliation{Department of Physics, Massachusetts Institute of Technology, Cambridge, MA, USA}

 \author{F.~Parmigiani}
 \affiliation{Elettra - Sincrotrone Trieste S.C.p.A., Strada Statale 14, km 163.5, Trieste, Italy}%Lines break automatically or can be forced with \\
 \affiliation{International Faculty, University of Cologne, Albertus-Magnus-Platz, 50923 Cologne, Germany}

\author{F.~Cilento}%
\email{federico.cilento@elettra.eu}
\affiliation{Elettra - Sincrotrone Trieste S.C.p.A., Strada Statale 14, km 163.5, Trieste, Italy}%

\begin{abstract}

CoSn is a prototypical kagome compound showing lattice-born flat bands with suppressed bandwidth over large parts of the Brillouin zone. Here, by means of time- and angle-resolved photoelectron spectroscopy, we provide direct evidence of the response to photoexcitation of the flat bands, that underlie information about localization in real space. In particular, we detect a sudden shift and broadening of the flat bands, while after one picosecond only a broadening survives. We ascribe both these effects to an ultrafast disruption of electron localization, which renormalizes the effective electron-electron interaction and affects the flat band dispersion. Since both variations are in the order of a few meV, our measurements suggest that the flat bands are resilient to near-infrared photoexcitation.

\end{abstract}

\maketitle

Kagome systems have recently attracted increasing attention owing to their capability of simultaneously hosting a variety of exotic phases of both correlated and topological nature \cite{Kang2019, Yin2019, Hu2022, Mazin2014, Liu2014}. Notable examples include high-temperature superconductivity \cite{Iglovikov2014, Ko2009, Jiang2022}, charge-density wave \cite{Neupert2021, Tan2021, Han2022} Bose-Einstein condensation \cite{Huber2010}, topological insulators \cite{Guo2009, Bolens2019} and anomalous Hall effect \cite{Neupert2011, Xu2015, Nakatsuji2015}. Several types of magnetic ordering have also been observed in coexistence with one or more of these phenomena \cite{Lin2018, kuroda2017evidence, Mielke1992}. The key feature of kagome systems is the presence of flat bands in the band structure, which originate from electron localization. Charge confinement descends from the peculiar geometry of the kagome lattice, whose eigenfunctions are compact localized states enabled by the destructive interference of nearest-neighbor electronic hopping pathways \cite{Sutherland1986, Leykam2018}. Additionally, when strong spin-orbit coupling is present, the band structure can display topologically nontrivial features \cite{Titvinidze2021, Liu2014}. Since the localization relies on quantum interference \cite{Leykam2018}, external perturbations may affect the stability of the electronic structure. The effect of a temperature increase in the band structure was theoretically predicted to be minimal up to 1000\,K \cite{Wan2022}. On the other hand, nonthermal excitation pathways are a promising tool to investigate correlated systems \cite{Giannetti2016}. Time-resolved optical spectroscopy has been applied to study the kagome metal $\mathrm{CsV_3Sb_5}$, where the pump pulse induces a nonthermal melting of its unconventional charge-density wave phase \cite{Wang2021}. At present, however, no report discusses the robustness of the electron localization of kagome-derived states upon ultrafast photoexcitation. Given its origin, it is reasonable to expect that the effective localization is altered by the perturbation of the electronic subsystem if energy is injected in it by means of a laser pulse. The resulting pump-induced modifications of the flat bands can be regarded as a fingerprint of the response in terms of charge localization. For this reason, time-and angle-resolved photoelectron spectroscopy (TR-ARPES) constitutes the ideal tool since it allows accessing the evolution of the band dispersion as the system relaxes from the photoexcited state to equilibrium \cite{Smallwood2016}. Among the recently discovered kagome compounds, the realization of a flat band extending over the entire Brillouin zone has proven challenging due to the presence of spurious effects such as defects or strong interlayer coupling \cite{Liu2018,Lin2018}, or because it lies above the Fermi level \cite{Kang2019}. CoSn is the first known case of a kagome metal in which the flat bands extend over the whole Brillouin zone, lying below and in the vicinity of the Fermi level \cite{kang2020topological, Sales2021}. The flat bands are thus accessible via direct photoemission and their response to ultrafast excitation with a laser pulse can be observed. We report time- and angle-resolved photoelectron spectroscopy (TR-ARPES) measurements on the kagome metal CoSn to investigate the robustness of a lattice-born flat band to optical excitation. 

\begin{figure*}[th!]
\centering
\includegraphics[width=0.9\textwidth]{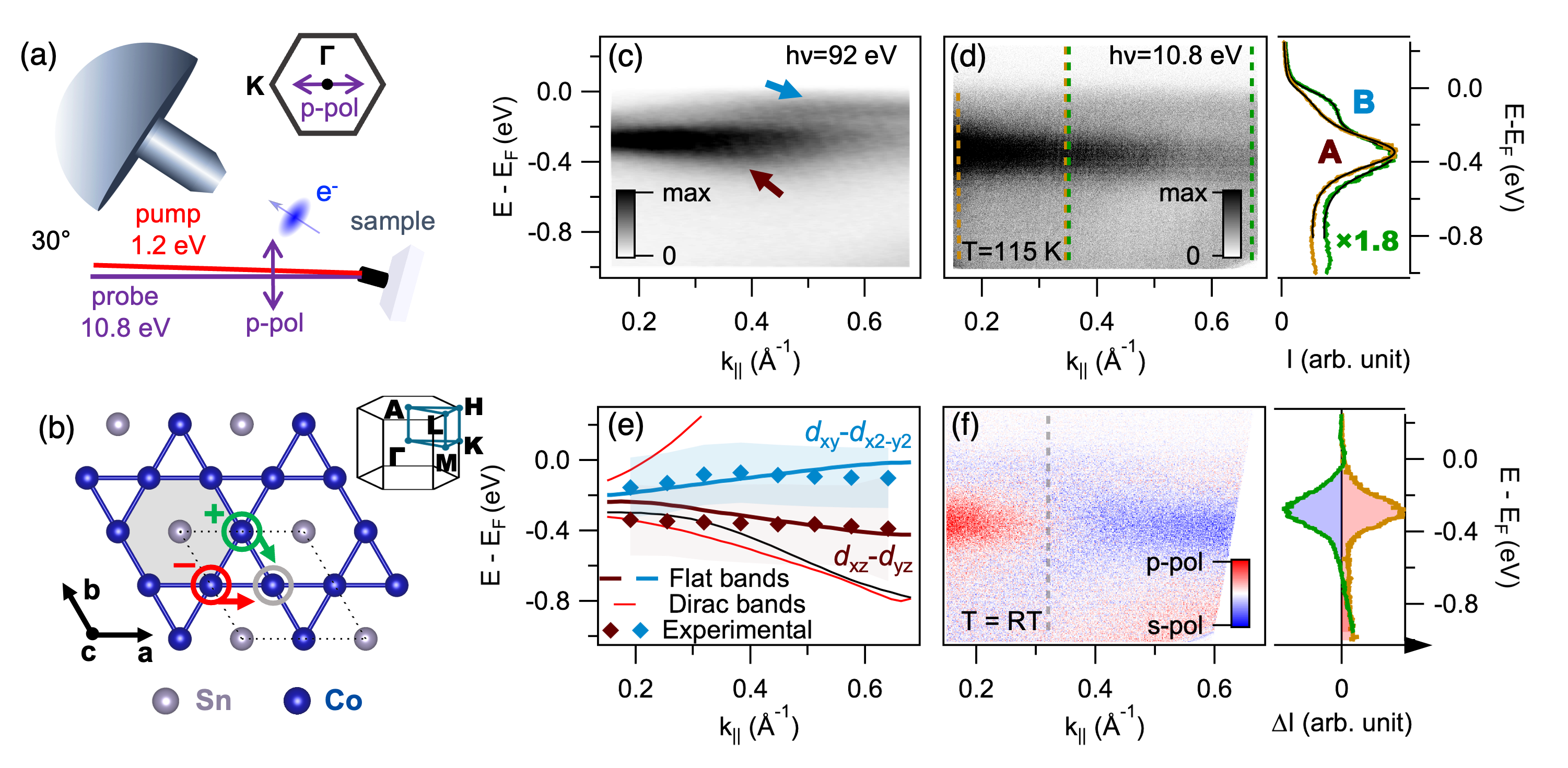}
\caption{
(a) Scheme of the experimental geometry (top view). Inset: relative orientation of the crystal Brillouin zone and the direction of the $p$ polarization in the geometry of the experiment. 
(b) Schematic lattice of a CoSn kagome layer. Colored circles denote the adjacent sites with opposite-sign phase of the wavefunction (green and red), causing hopping to the nearest-neighbor (grey) to vanish. The grey-shaded area indicates the real space region of electron confinement. Inset: first Brillouin zone of CoSn. 
(c) Band structure of CoSn along the $\mathrm{\Gamma}$K path measured with 92\,eV photons. The blue (brown) arrow indicates the upper (lower) flat band \cite{kang2020topological}. 
(d) Dispersion measured with $p$-polarized photons at 10.8\,eV and 115\,K along $\mathrm{\Gamma-K}$ ($\mathrm{A-H}$). In the right panel, EDCs extracted in the momentum regions of corresponding color: orange for small momentum values, green for large momentum values. The best fit is overlaid as a black line. A and B identify two Lorentzian peaks used to fit the EDCs. 
(e) DFT calculations of the band structure of CoSn along the $\mathrm{\Gamma-K}$ direction in the vicinity of the Fermi level. The upper and lower flat bands are indicated with blue and brown lines respectively, the Dirac band in red and a non-kagome derived band in black. The degrees of freedom of the $3d$ orbitals from which the two bands originate are also reported. The square markers indicate the position of the two peaks resulting from the fit of the EDCs integrated in the momentum region centred at the marker position, with the Lorentzian width indicated as a shaded area. 
(f) Photoemission intensity difference between the maps acquired with $p$- and $s$-polarized photons, both acquired at room temperature. The momentum value at which the difference crosses zero is marked as a gray line.  
}
\label{fig:phys}
\end{figure*}

The measurements have been performed at the \mbox{T-ReX} laboratory (Elettra, Trieste), taking advantage of a novel ultrafast source of photons at 10.8\,eV. The setup is based on the output of an Yb-doped fiber laser (Coherent Monaco), a fraction of which is used as a pump (1.2\,eV). The remainder is used to generate the ninth harmonics at 10.8\,eV through a cascade of frequency upconversion stages in BBO crystals and in Xe gas. Details on the harmonic generation setup are reported elsewhere \cite{Peli2020}. For the present measurements the repetition rate of the laser was set to 200\,kHz and the pump fluence to 260\,$\mathrm{\mu J/cm^2}$. Fig.~\ref{fig:phys}(a) shows the top view of the endstation, indicating the relative orientation of the beams, the sample and the analyzer. The orientation of $p$ polarization with respect to the experimental geometry and the Brillouin zone (inset) is also shown. The CoSn crystals grow along the $c$ axis as thin posts of hexagonal section and diameter up to 0.8\,mm. The crystal orientation has thus been determined by \textit{ex-situ} inspection of the sample surface. This method has an estimated uncertainty of about $5^{\circ}$. Crystals were cleaved \emph{in situ} by fracture and measured at a pressure of $10^{-10}$\,mbar at a base temperature of 115\,K, unless otherwise specified. We stress that the size and shape of CoSn makes the experiment challenging. In particular, fringing fields from the sample edges contribute to alter the energy and momentum distribution of photoelectrons, with a larger impact at smaller kinetic energies. The photon energy of 10.8\,eV is thus crucial to mitigate these effects, which can significantly disrupt the quality of measurements at photon energies of $\sim 6$\,eV. 

We used the Vienna Ab initio Simulation Package \cite{Kresse1996,Kresse1996a} to perform the density functional theory \cite{Bloechl1994} calculations of the CoSn band structure, already reported in \cite{kang2020topological}. The Projector Augmented Wave \cite{Perdew1996} type pseudo-potentials and the exchange-correlation energy functional parameterized by Perdew, Burke, and Ernzerhof(PBE) were employed in the calculations. To converge the electronic ground state, we have used a 350 eV energy cut-off, a 15 x 15 x 11 momentum space sampling grid with spin-orbit coupling corrections. 

CoSn crystals are comprised of two-dimensional kagome layers (Fig.~\ref{fig:phys}(b)) intercalated along the $c$ axis with honeycomb layers of Sn. The Co atoms (blue) are arranged in corner-sharing triangles and constitute the kagome lattice. This structure leaves empty hexagonal sites (grey-shaded in the figure) which are occupied by Sn atoms. The linear superposition of wavefunctions of opposite-sign phases centered at the kagome sites (red- and green-circled atoms) is an exact eigenstate of a tight-binding model Hamiltonian including only nearest-neighbor hopping. Such state is localized in the grey-shaded hexagon, as destructive interference at the nearest neighbor (grey-circled atom) causes the hopping to this site to vanish \cite{Sutherland1986, Leykam2018}. In CoSn, these states arise from the $3d$ orbitals of Co. From the energy-scale point of view, the itinerancy of charges is quenched due to the confinement, and the potential energy due to Coulomb repulsion dominates over the kinetic energy. This scenario is typical of correlated systems and leads to the emergence of many-body physics \cite{Damascelli2003, Yanagisawa2021}. CoSn displays a pair of flat bands, as it can be seen from the ARPES measurements in Fig.~\ref{fig:phys}(c) \cite{kang2020topological}. The two arise from different degrees of freedom of the $3d$ orbitals, namely the upper one from the $d_{xy}/d_{x^2-y^2}$ and the lower one from the $d_{xz}/d_{yz}$ components. The small dispersion along the $\mathrm{\Gamma-K}$ direction is a consequence of nonzero next-nearest-neighbor hopping term \cite{Bolens2019}. In addition to the flat bands, the band structure of CoSn hosts a pair of Dirac bands, which display a spin-orbit coupling gap at the touching point with the flat band, endowing the latter with nontrivial topological character \cite{kang2020topological}. 

The equilibrium dispersion measured along the $\mathrm{\Gamma-K}$ ($\mathrm{A-H}$) direction with 10.8\,eV photons is displayed in Fig.~\ref{fig:phys}(d). The strong feature at \mbox{-0.3\,eV} is compatible with the observation of the lower flat band in Fig.~\ref{fig:phys}(c) (brown arrow). The energy distribution curves (EDCs) integrated in the momentum regions of corresponding color are displayed in the right part of the panel. Both curves suggest the presence of (at least) two bands. Hence, to reproduce the EDCs we multiply a sum of Lorentzian peaks by the Fermi-Dirac distribution: 

\begin{eqnarray}
I(E) = \frac{1}{e^{(E-E_F)/k_BT_e}+1} \cdot \sum_{i} a_i L_i
\label{eq1},
\end{eqnarray}

where $E_F$ is the Fermi energy, $T_e$ the electronic temperature, and $L_i$ is a Lorentzian peak normalized to 1, so that $a_i$ is the area under $i$-th peak. This function is then convoluted with a Gaussian to account for the experimental resolution of 70\,meV FWHM. A sigmoid background was used and an additional peak is required to mimic the intensity at higher binding energies. In the range of interest only the tail of this peak contributes. We tracked the dispersion of these two features by extracting EDC curves integrated in eight momentum regions across the whole measured energy \emph{vs.} momentum map, and compared the results with the DFT predictions of the dispersion along $\mathrm{\Gamma-K}$ (Fig.~\ref{fig:phys}(e)). The peak positions are displayed as squares centered in the integration region, with the width of the Lorentzian peak indicated as a shaded area. As the plot shows, the two features have a flat dispersion at large $k$ values, and get closer towards $\mathrm{\Gamma}$ due to a small dispersion. This comparison allows assigning the two features to the upper and lower flat bands, lying $\sim -0.13$\,eV and $\sim -0.37$\,eV below the Fermi level. At 10.8\,eV photon energy, the value of $k_z$ lies at half distance between $\Gamma$ and A, hence the observed discrepancies might be due to the small dispersion of the flat bands along $k_z$ \cite{kang2020topological}. According to our assignment, photoemission at 10.8\,eV photon energy prevents observation of the Dirac and the non-kagome bands. These features remain undetectable also by changing the probe polarization. We also detect a probe linear dichroism on peak A, shown in Fig.~\ref{fig:phys}(f). The momentum value of zero-crossing of the dichroic signal on peak A (grey dashed line) is within experimental uncertainty the same at which the upper flat band changes its dispersion.

\begin{figure}
\centering
\includegraphics[width=\columnwidth]{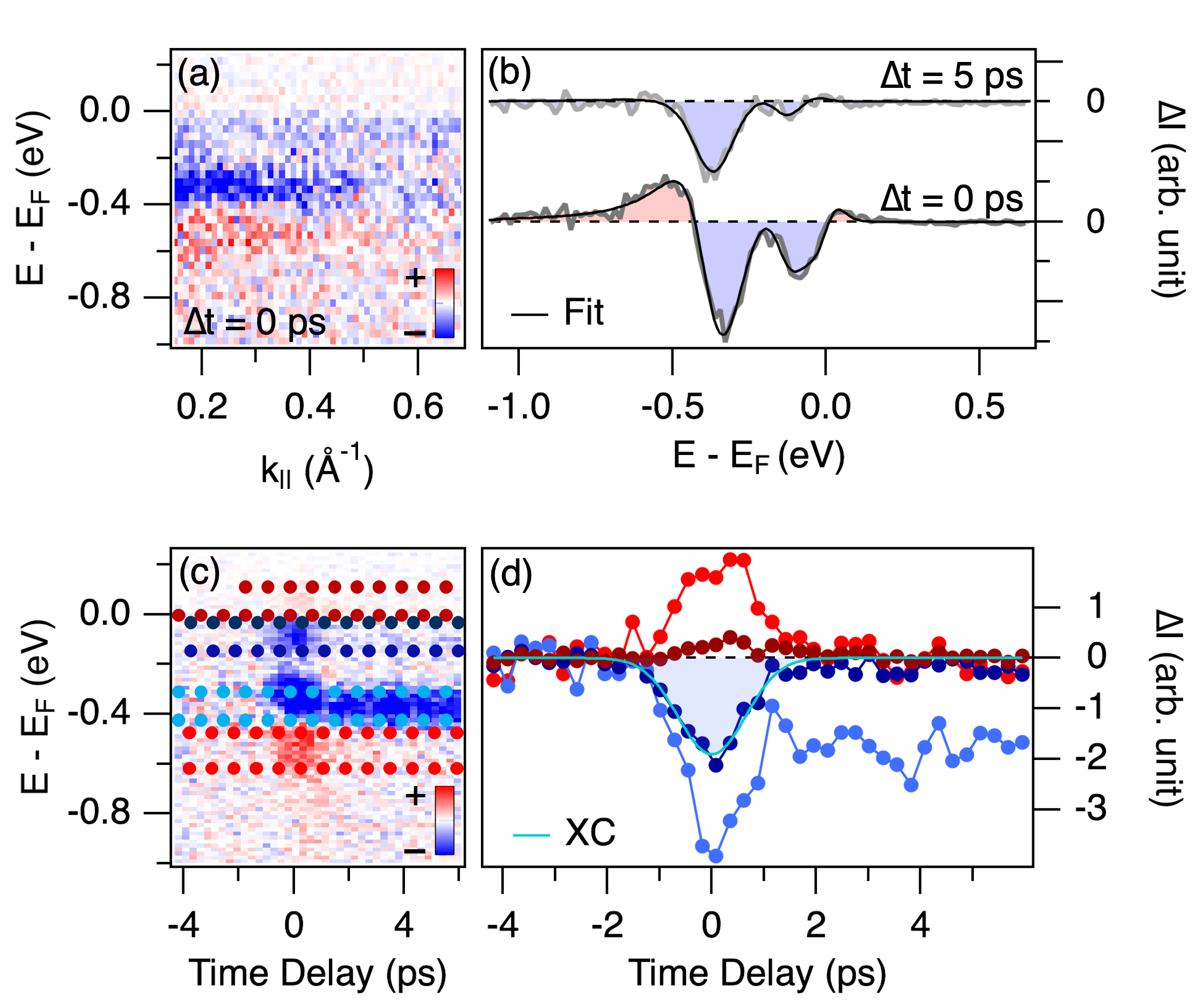}
\caption{
(a) Difference between the map at time zero and the average before time zero.
(b) Differential EDCs integrated in the whole momentum region region at 0 and 5 ps time delay. The black superimposed line is the fit, described in the main text. 
(c) Differential EDCs plotted in a color map as a function of time delay.
(d) Time traces extracted in the energy region marked by the dotted lines in (c). The light-blue trace is the gaussian fit to the dark-blue curve.
}
\label{fig:dyns}
\end{figure}

The effect of photoexcitation on the measured band structure is shown in Fig.~\ref{fig:dyns}. To highlight the changes induced by the pump pulse, the map acquired at negative time delays is subtracted from the one acquired at time zero, and displayed in panel (a). By virtue of the negligible dispersion of the flat bands discussed above, we can integrate the map in the whole momentum range. This produces the differential EDC curves (dEDCs) shown in panel (b) for two selected time delays (0\,ps and 5\,ps). In correspondence of the two flat bands we mostly observe a decrease in the photoemission intensity, although a net increase appears around -0.5\,eV, close to the bottom of the lower flat band. A small increase is also observed above the Fermi level. Time traces are extracted in four different energy regions and displayed in (d). At all the energies, the maximal variation is detected at time zero. The intensity increases both above the Fermi level and below the lower flat band. The intensity dynamics extracted just below $E_F$ can be fit with a Gaussian whose width is comparable to the time resolution of the setup of 1.3\,ps. Conversely, the negative signal related to the lower flat band, after an intense dip at time zero, reaches a plateau that lasts more than the investigated delay range of $\sim$\,6\,ps. The same effect, although with a smaller nonequilibrium signal, occurs for the upper flat band. Interestingly, the negative signal in the region of the lower flat band shifts downwards in energy after the cross correlation of pump and probe pulses.

\begin{figure}
\centering
\includegraphics[width=\columnwidth]{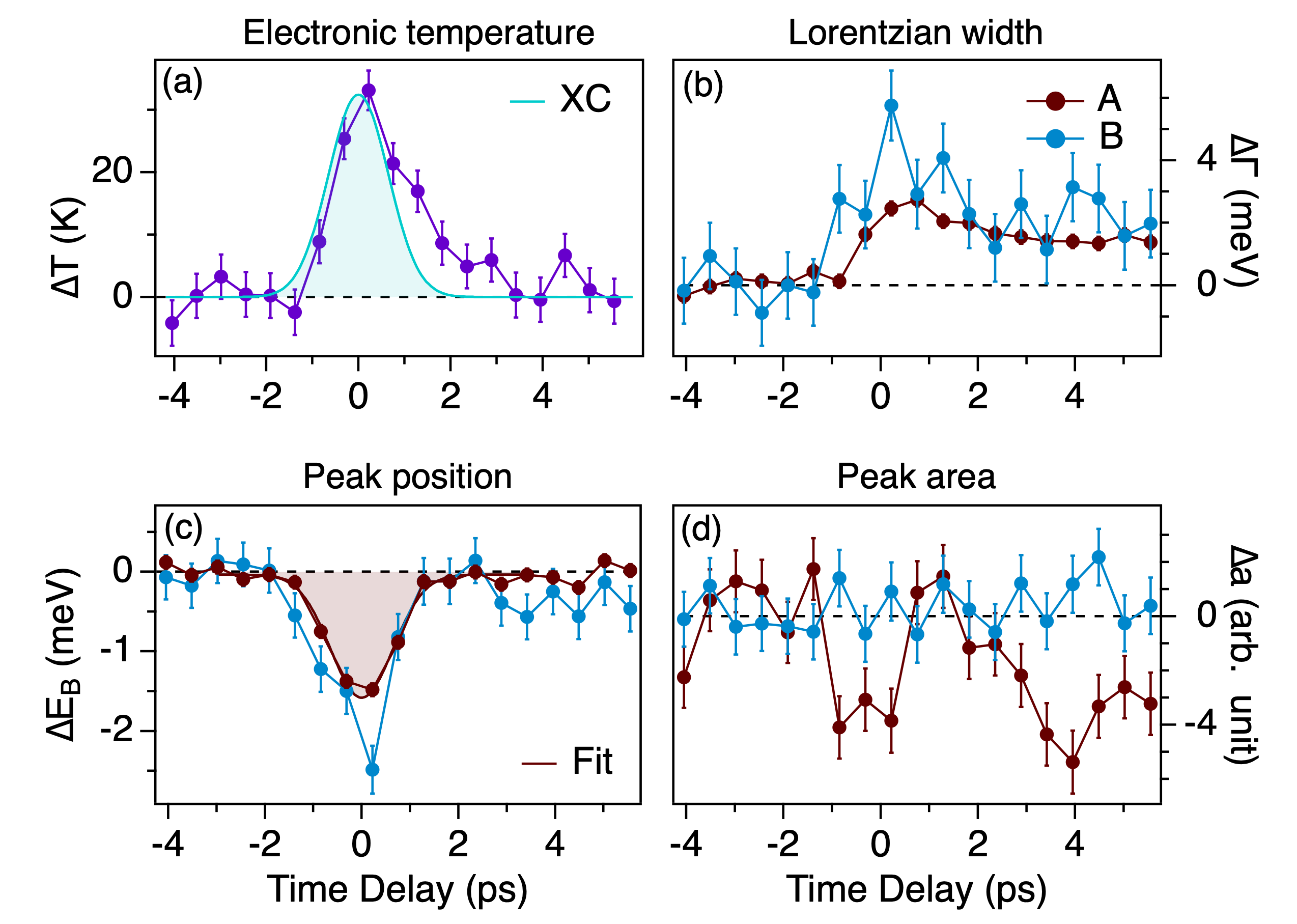}
\caption{
Variation with respect to equilibrium of the parameters extracted by fitting the dEDCs, as a function of time delay. Bars indicate the uncertainty of the fitting procedure on the parameters. (a) Electronic temperature. The solid line is the gaussian fit reported from Fig.~\ref{fig:dyns}(d). (b) Lorentzian width of peak A (brown) and B (light blue). (c) Position (binding energy) of peaks A and B. The solid line is the best fit of the A position dynamics to a Gaussian. (d) Area under each Lorentzian peak. 
}
\label{fig:fitdyns}
\end{figure}

To identify which lineshape modifications underlie these spectral changes, we fit each dEDC as a function of the time delay. The average over negative time delays is fit with the same model described for the equilibrium data. This is subtracted to a function of identical form but with modified parameters, to obtain the fitting function for the dEDCs. To fit the data, the electronic temperature in the Fermi-Dirac distribution and the parameters of the two flat bands (area, position and width) are kept free. The uncertainty on the fit parameters is extracted propagating in the fitting procedure the Poissonian variance on the number of counts  in each energy bin. The quality of fit to the dEDCs is evident on the selected traces shown in Fig.~\ref{fig:dyns}(b), while the values of the parameters of best fit are plotted as a function of time delay in Fig.~\ref{fig:fitdyns}(a-d). The first important observation is that the A and B peaks display markedly similar dynamics, compatibly with the common origin from the $3d$ orbitals of Co. All the parameters, except for the peak area, display a larger change at time zero. The observed spectral changes are thus the result of an increase in the electronic temperature, along with a sudden shift and broadening of the flat features which explain the observation of an increase in photoemission intensity at -0.5\,eV below the Fermi level in Fig.~\ref{fig:dyns}. The subsequent relaxation dynamics are qualitatively distinct for the different fitting parameters. The comparison of the temperature evolution with the gaussian fit found in Fig.~\ref{fig:dyns} shows the presence of dynamics beyond the cross-correlation. The width of the flat features instead sets to a plateau of $\sim$\,2\,meV, $\sim 1 \%$ higher than equilibrium, lasting longer than 6\,ps. Finally, the agreement of the peak position relaxation with a gaussian of width comparable to the experimental cross-correlation (brown solid line in panel (c)) indicates that the observation of its dynamics is limited by our experimental resolution. The peak areas, instead, show no clear dynamics across time zero, signaling that charge redistribution from or to the flat band is negligible. These observations point towards the broadening of the flat bands as the origin of the long-lasting nonequilibrium signal visible in Fig.~\ref{fig:dyns}(c),(d). 

We interpret our results assuming that the ultrafast pump pulse triggers an electron delocalization, and hence affects the effective correlation strength. Both on the short and on the long timescales, the energy injected by the pump pulse causes an increase in the electronic temperature, indicative of an increased kinetic energy of the charge carriers. The shift and broadening of the flat bands can be ascribed to reduced effective electron correlations. Indeed, a renormalization of the correlations can cause band shifts in the spectrum. Moreover, delocalization is expected to increase the flat band dispersion. Since we integrate on a wide momentum region, and due to our energy resolution, a small increase in dispersion is seen as a broadening. The variation of the lineshape parameters is not only larger at time zero than at later delays, but also qualitatively different, as the shift of the two peaks is recovered after the cross-correlation. We conclude that the presence of the pump induces a disruption of the electron localization affecting both the position and the width of the bands. This stronger renormalization has a short lifetime and vanishes as soon as the pump leaves the sample, signaling the recovery of the electron localization. That the broadening is not purely thermal is guaranteed by the concomitant shift in binding energy. The case is different for the later time delays, where only the small broadening survives. Complete recovery could be hindered by some bottleneck effect that slows down the recovery of the electronic temperature and also preserves the itinerancy of a smaller fraction of charges, which are sufficient to induce a band broadening but not a shift. We note that the order of magnitude of this effect is compatible with a thermal broadening due to a lattice temperature few tens of K higher than equilibrium. Nonetheless, the small entity of lineshape changes in the flat bands points to a robustness of the localization mechanism upon infrared photoexcitation. 

In conclusion, we report on the first time-resolved photoemission study on the ultrafast dynamics of flat bands in a kagome system. The measurements show that the flat bands of CoSn undergo a small but detectable modification upon photoexcitation. We ascribe these effects to electron delocalization triggered by the ultrashort pump pulse, which reduces the effective electron-electron interaction and renormalizes the flat band dispersion. The majority of charges is relocalized within one picosecond, while a broadening of the flat bands lasts more than 6\,ps, signaling either a partial persistent delocalization or an elevated lattice temperature. The small extent of both effects allows us to conclude that the flat bands are robust against infrared photoexcitation. More generally, the present work calls for a deeper understanding of the response of a system in which electronic correlations are enhanced by localization, and paves the way for the exploration of the nonequilibrium properties of kagome systems. 

\begin{acknowledgments}
This work was funded, in part, by the Gordon and Betty Moore Foundation EPiQS Initiative, Grant No. GBMF9070 to J.G.C (computation), NSF grant DMR-1554891 (material analysis), and ARO Grant No. W911NF-16-1-0034 (technique development). P.M.N. acknowledges support by the STC Center for Integrated Quantum Materials, NSF grant number DMR-1231319.
\end{acknowledgments}

\providecommand{\noopsort}[1]{}\providecommand{\singleletter}[1]{#1}%

\end{document}